\documentclass[global,twocolumn]{svjour}
\usepackage{graphics}
\journalname{Applied Physics B}
\begin{document}
\title{Occurrence control of charged exciton for a single CdSe quantum dot at cryogenic temperatures on an optical nanofiber}
\author{K. Muhammed Shafi\inst{1} \and Kazunori Iida\inst{2}\and Emi Tsutsumi\inst{2}\and Akiharu Miyanaga\inst{2} \and Kohzo Hakuta\inst{1}
\thanks{\emph{Corresponding author:} k.hakuta@cpi.uec.ac.jp}%
}                     
%
%
\institute{Center for Photonic Innovations and Institute for Laser Science, University of Electro-Communications, Chofu, Tokyo 182-8585, Japan. \and NS Materials Inc., Tsukushino, Fukuoka 818-0042, Japan.}

\date{Received: date / Revised version: date}
%
\maketitle
\begin{abstract}
We discuss photo-luminescence characteristics of CdSe core/shell quantum dots at cryogenic temperatures using a hybrid system of a single quantum dot and an optical nanofiber. The key point is to control the emission species of quantum dot to charged excitons, known as trions, which have superior characteristics to neutral excitons. We investigate the photocharging behavior for the quantum dots by varying the wavelength and intensity of irradiating laser light, and establish a method to create a permanently charged situation which lasts as long as the cryogenic temperature is maintained. The present photocharging method may open a new route to applying the CdSe quantum dots in quantum photonics, and the hybrid system of photocharged quantum-dot and optical nanofiber may readily be applicable to a fiber-in-line single-photon generator.
\end{abstract}

\section{Introduction}
Recent progress in nanophotonics has opened new prospects in the field of quantum information technologies. A central issue of nanophotonics is to develop hybrid systems for a single quantum emitter and nano-waveguides so that single photons can be generated or manipulated in fiber networks \cite{EmiReview,Nayak1,Sipahigil}. Regarding quantum emitters, various solid state emitters have been investigated so far \cite{EmiReview,Skoff,Yalla1,Fujiwara,Schell1}. Among them, one promising candidate is excitons of CdSe core/shell quantum dots (QDs) which provide high quantum yield (QY) and tunability of emission wavelength \cite{PietrygaReview,David}. Systematic spectroscopic studies have been performed for QDs from room temperature to cryogenic temperatures, and it has been recognized that the QD emission occurs not only in neutral form but also in charged form \cite{Louyer}. Excitons for charged QDs are known as trions. Although excitons for neutral QDs show nearly unity QY from room temperature to cryogenic temperatures, they have one weak point from a viewpoint of quantum photonic applications. Since the exciton upper state consists of two levels with a metastable lower level, the so called dark state, the effective decay time of the exciton becomes rather long and the system response using neutral excitons becomes slow \cite{Biadala,Labeau}. On the other hand, the upper state of charged excitons is effectively one level and the decay time is one-order faster than that for neutral excitons \cite{Javaux}. Regarding the QY of charged excitons, it is rather low at room temperatures, but it has been reported that the QY becomes higher as the temperature is lowered and reaches nearly unity at cryogenic temperatures using thick-shell QDs with CdS or CdS/ZnS shell \cite{Javaux,Fernee2}. Thus for quantum photonic applications, charged excitons are superior to neutral excitons at cryogenic temperatures. One may wonder about the cryogenic condition from the viewpoint of applications. However, in the field of quantum photonics, cryogenic technologies are becoming common technology especially for single photon detectors \cite{Natarajan}, and various new technologies operated at cryogenic temperatures have been developed \cite{Sipahigil,Bhaskar,Schlehahn}. Therefore, it would be valuable to explore characteristics of charged excitons at cryogenic temperatures for quantum photonics. Hereafter we refer to neutral excitons simply as excitons and charged excitons as trions.

So far, many groups have discussed the characteristics of excitons and trions for CdSe QDs at cryogenic temperatures \cite{Liu,Fernee,Gomez}. Regarding the occurrence of neutral and charged conditions, both conditions occur almost randomly for most cases, and controlling the QD emission to achieve trion emission is an important issue from the viewpoint of quantum photonic applications. Javaux {\it{et al.}} \cite{Javaux} reported that several in hundreds of QDs behaved as charged QDs. Rinehart {\it{et al.}} \cite{Rinehart} discussed photochemical electron doping for QDs in colloidal solution to create negatively charged QDs. However, to extend the photonic applications, it would be beneficial to control the trion occurrence with physical techniques. Regarding this point, two groups observed photocharging for QDs after they were irradiated by laser light \cite{Javaux,Fernee2}, but the process has not been investigated systematically.

In this paper we report how the occurrence of trion emission in a single CdSe QD can be controlled optically at cryogenic temperatures by irradiating with laser light. We carry out experiments using a hybrid system of a tapered optical fiber with sub-wavelength waist diameter, an optical nanofiber (ONF) and a single QD, which we reported previously \cite{Shafi} in the context of quantum photonic applications. We extend the measurements to systematically investigate photoluminescence (PL) characteristics of the QDs by varying the wavelength and intensity of the irradiation laser light. We find that the degree of photocharging can be controlled with irradiation wavelength and intensity. Furthermore, we have found that the irradiation at 355 nm can lead QDs to a permanently photocharged situation, which lasts as long as the cryogenic condition is maintained. We find also that the charging can be removed by raising the temperature to 150 K. We discuss a possible mechanism of the observed photocharging. The present results may trigger research works into more detailed understanding of the photocharging mechanism, and may also open a new way to apply trion-QDs in quantum nanophotonics at cryogenic temperatures.

\begin{figure}[t!]
\centering
\resizebox{0.5\textwidth}{!}{%
  \includegraphics{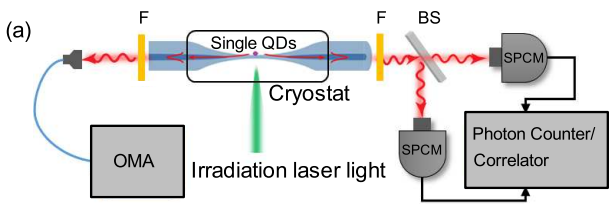}
}
\caption{Measurement setup for the QD/ONF hybrid system: OMA, F, BS, and SPCM denote optical multichannel spectrum analyzer, color-glass filter, beam splitter, and single-photon counting module, respectively.}
\label{Fig1}
\end{figure}

\section{Experimental system}

We used CdSe QDs with a gradient thick shell with an outermost ZnS shell, which was synthesized using a procedure described previously \cite{Shafi}. A brief sketch of the experimental arrangement is displayed in Fig. \ref{Fig1}.  We used ONFs with a waist diameter of 310 nm, for which PL photons around 640 nm are channeled into the nanofiber guided modes with an efficiency of 22\% \cite{Kien1,Yalla2}. We deposited several single QDs on the waist of an ONF with a separation of 150 $\mu$m using a computer-controlled sub-picolitre dispenser system \cite{Yalla1,Shafi}, and each single QD was irradiated with a laser light perpendicularly to the ONF with a focused spot radius of 15 $\mu$m. The polarization of the irradiation lasers was linear, and the axis was chosen to be perpendicular to the nanofiber axis so that the PL photons into the ONF guided mode led to maximum photon counts. PL spectrum, photon counts, and photon correlations were measured simultaneously through both ends of the fiber after passing through color-glass filters (O56, HOYA) to remove scattered laser light. We also measured the polarization characteristics for the PL photons by inserting a polarizer (analyzer) together with the filters.

The system of QD/ONF was installed into a custom designed cryostat which cooled the system down to 3.7 K via He buffer gas cooling. One specific feature of the cryostat system is that the optical transmission of the ONF can keep its room temperature value at cryogenic temperatures with a negligible drop of only 0.5\%. Regarding the PL spectrum measurements, four cw-lasers at 532, 455, 405, and 355 nm were used for irradiation. Irradiation intensity dependence was systematically measured at each wavelength. We used fresh QDs for a series of measurements at each irradiation wavelength. PL spectra were measured with a spectral resolution of 0.3 nm (1.0 meV) using an optical multichannel analyzer (MS3504i, SOL instruments) equipped with a water-cooled CCD camera (DV420A-OE, Andor).
For PL measurements, we integrated the signal for a measurement time of 120 s or longer.
Photon correlation measurements were performed using two single photon counting modules (SPCM-AQRFC, Perkin Elmer) and a time-correlated single photon counting system (Picoharp 300, PicoQuant).
Radiative decay behavior was measured using a picosecond pulsed-laser at 532 nm (FP200-SH, Ekspla) with a pulse width of 20 ps FWHM and a repetition rate of 500 kHz, by observing temporal correlations between the laser pulse and PL photons. The laser pulse energy was fixed to a low value of 0.2 nJ/pulse so that 
the average number of photons absorbed per pulse per QD is less than one. We calculated the pulse energy by estimating the QD photoabsorption cross-section at 532 nm \cite{Leatherdale}.

\section{Characteristics at room temperature}

\begin{figure}[t!]
\centering
\resizebox{0.5\textwidth}{!}{%
  \includegraphics{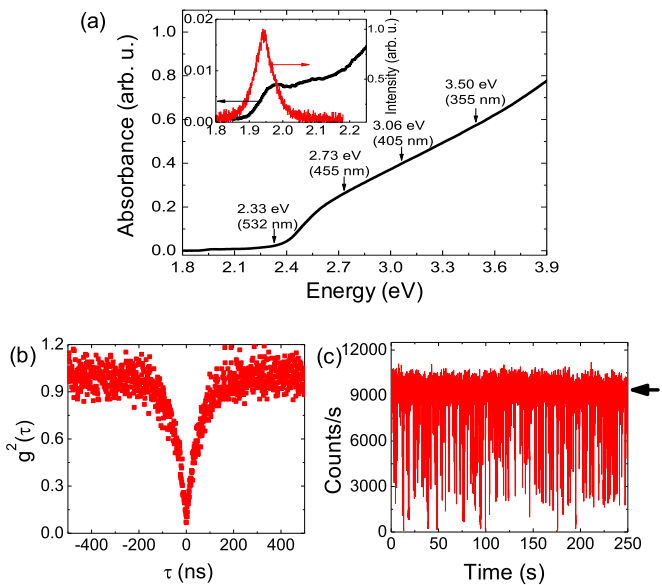}
}
\caption{QD characteristics at room temperature. (a) The absorbance spectrum for QDs in toluene solution measured with a conventional absorbance spectrometer. Laser photon energies used for excitation are indicated by arrows. In the inset, enlarged view around the band edge is shown, together with the PL spectrum for a single QD measured with the ONF setup by a red curve. (b) Photon correlation characteristics for a single QD on an ONF by irradiating 532 nm laser light at an intensity of 20 W/cm$^2$. (c) Photon counting behavior for PL photons from a QD on an ONF by irradiating 532 nm laser light at an intensity of 20 W/cm$^2$.}
\label{Fig2}
\end{figure}

Figure \ref{Fig2}(a) displays an absorbance spectrum for QDs in toluene solution at room temperature measured with a conventional absorbance spectrometer. In the inset, enlarged view around the band edge is shown, together with PL spectrum for a single QD measured with the ONF setup, showing a center wavelength of 640 nm (1.94 eV) and an FWHM of 20 nm (61 meV). In Figs. \ref{Fig2}(b, c), photon correlations and photon counting characteristics are displayed. The photon correlation characteristics reveal a single emitter behavior with an anti-bunching dip close to zero. The photon counting characteristics show a PL intermittency peculiar to a single QD at room temperature. Note that the high photon count level marked by an arrow corresponds to a QY of nearly unity, details of which were discussed previously \cite{Shafi}.

\section{Characteristics at cryogenic temperatures}

In the following, we summarize the results obtained for four different irradiation wavelengths at cryogenic temperatures. Measurement temperature was set at 3.7 K, except specially specified cases.

\begin{figure}[t!]
\centering
\resizebox{0.5\textwidth}{!}{%
  \includegraphics{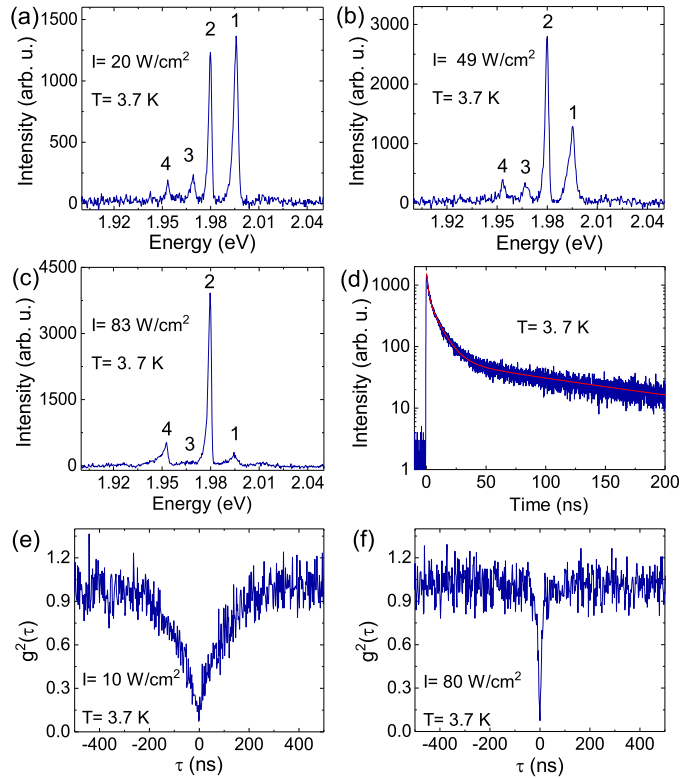}
}
\caption{PL characteristics at 3.7 K observed for 532 nm irradiation. (a-c) PL spectrum for a single QD obtained with three different irradiation intensities, 20, 49, and 83 W/cm$^2$. (d) PL decay profile of a single QD measured with a ps-pulsed laser at 532 nm plotted in semi-log format. Fitted curve is drawn by a red curve. (e, f) Photon correlation signals observed at irradiation intensities of 10 and 80 W/cm$^2$.}
\label{Fig3}
\end{figure}

\begin{figure}[t!]
\centering
\resizebox{0.5\textwidth}{!}{%
  \includegraphics{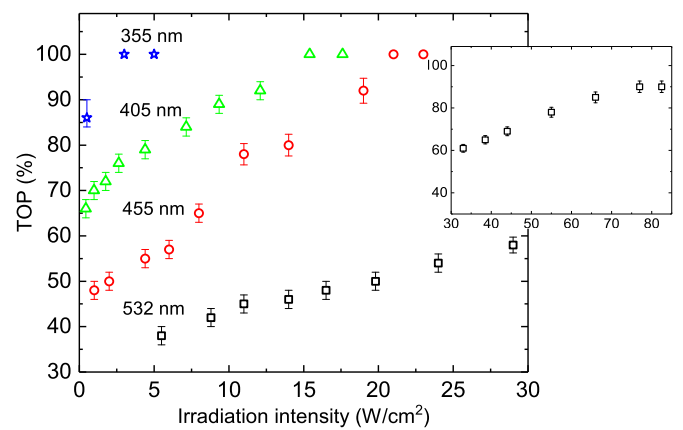}
}
\caption{TOP versus irradiation laser intensity measured at 3.7 K. For the irradiation, four lasers at wavelengths of 532, 455, 405, and 355 nm were used, and corresponding results are marked by black squares, red circles, green triangles, and blue asterisks, respectively. The inset shows a plot for 532 nm irradiation for intensities higher than 30 W/cm$^2$. }
\label{Fig4}
\end{figure}

\subsection{532 nm irradiation}
PL spectra excited at 532 nm were measured for irradiation intensities from 5 to 83 W/cm$^2$, and spectra at 20, 49, and 83 W/cm$^2$ with an integration time of 120 s are displayed in Figs. \ref{Fig3}(a-c). Peaks 1 and 2 were assigned to exciton and trion, respectively, based on the previously reported results \cite{Louyer,Biadala}. Peaks 3 and 4 are the LO-phonon sideband for exciton and trion, respectively. The spectral width of the exciton was broader than that of the trion, since the exciton upper state consists of two levels, bright and dark state, with a splitting of 1.5 meV, which were not resolved. It is clearly seen that the occurrence of the trion increases with increasing irradiation intensity. Note that the occurrence of the exciton and trion reached a stationary condition for the integration time of 120 s as described previously \cite{Shafi}. We defined the trion occurrence probability (TOP) as a ratio between the photon number emitted by the trion and the total photon number emitted by the trion and the exciton under the stationary condition, which can be estimated by integrating the area of peak 2 and the whole area for peak 1 and 2.  We fitted the observed spectral profiles by Lorentzian profile to calculate the area. Obtained TOP values are plotted versus irradiation intensity in Fig. \ref{Fig4} by black squares. Data points for intensities higher than 30 W/cm$^2$ are shown in the inset. It is clearly seen that the TOP value increases monotonically from 37\% at 5 W/cm$^2$ with increasing irradiation intensity and saturates at 77 W/cm$^2$ with a value of 90\%. After reaching the saturation, we remeasured the spectra by lowering the irradiation intensity, and observed the same spectrum and TOP value at each intensity. The observations mean that the strong 532 nm irradiation realizing the TOP saturation has not led to any irreversible change for the QD. Regarding photon counting behavior at 3.7 K, the intermittency observed at room temperature was completely removed and showed the same high photon count level as that at room temperature. This situation is the same as that discussed in refs. \cite{Shafi,Javaux}, leading to the fact that the trion QY was improved to the same QY as the exciton at room temperature. Figure \ref{Fig3}(d) displays the observed decay profile, which was well fitted by a sum of three exponential decay curves with decay times of 1.47$\pm$0.05, 9.2$\pm$0.3, and 135$\pm$9 ns, as shown by a red curve. The decay time of 9.2 ns corresponds to the trion decay time, and decay times of 1.47 and 135 ns are assigned to fast and slow components of the exciton decay process, respectively \cite{Biadala,Labeau}. Figures \ref{Fig3}(e, f) display the photon correlation characteristics observed at irradiation intensities of 10 and 80 W/cm$^2$, respectively. Although the recovery time was quite different for the two characteristics, both characteristics clearly showed a deep anti-bunching dip close to zero, revealing a single quantum emitter condition. We estimated the effective recovery time by using a single exponential fitting to be 105$\pm$21 and 9$\pm$2 ns for the intensities of 10 and 80 W/cm$^2$, respectively, and the values can be understood through corresponding TOP values. For the 10 W/cm$^2$ irradiation, the TOP value is about 40\% (see Fig. \ref{Fig4}). This means that 60\% of emitted photons were from exciton which has a slow decay component due to the metastable dark state. On the other hand, for the 80 W/cm$^2$ irradiation, the corresponding TOP value is about 90\%, where 90\% of emitted photons were from trion which has a short decay time of around 10 ns. We should note that regarding the measurements for 455, 405, and 355 nm irradiation described below, photon correlations also resulted in a deep anti-bunching dip with a recovery time consistent with the corresponding TOP value. These results on photon-correlations lead to the fact that a single quantum emitter condition was maintained throughout all irradiation conditions for the present work.

\begin{figure}[t!]
\centering
\resizebox{0.5\textwidth}{!}{%
  \includegraphics{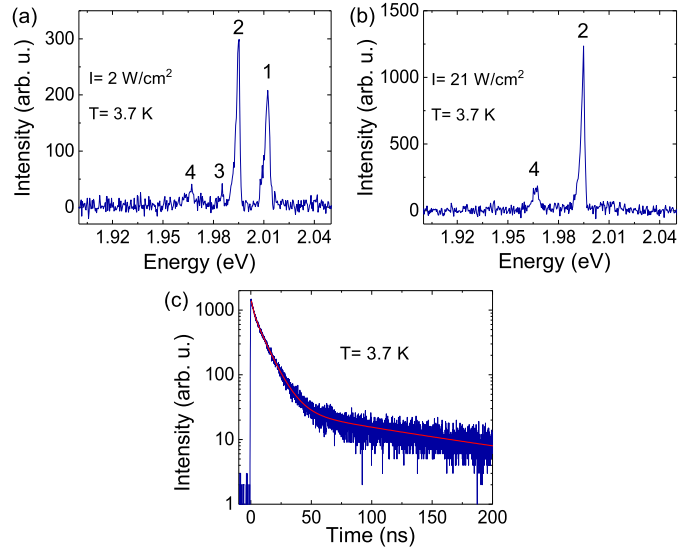}
}
\caption{PL characteristics observed for 455 nm irradiation at 3.7 K. (a, b) PL spectra measured for two different irradiation intensities, 2 and 21 W/cm$^2$. (c) PL decay profile plotted in semi-log format measured using a ps-pulsed laser at 532 nm after realizing the fully charged condition. The fitted curve is shown in red.}
\label{Fig5}
\end{figure}

\subsection{455, 405 nm irradiation}
PL spectra excited at 455 nm for two irradiation intensities with an integration time of 120 s are displayed in Figs. \ref{Fig5}(a, b), and TOP behavior versus irradiation intensity is plotted in Fig. \ref{Fig4} by red circles. We measured the PL spectra also with an integration time of 180 and 240 s, but did not find any change in the spectra, leading to the fact that the TOP reached a stationary condition with an integration time of 120 s. The TOP plot showed a similar behavior as that for 532 nm, but it is shifted to the lower intensity side. The TOP value increased with increasing irradiation intensity from 48\% at 2 W/cm$^2$ and a fully charged condition of 100\% TOP value was realized at 21 W/cm$^2$. Regarding the 455 nm irradiation process, we found that the strong irradiation induced an irreversible change in the QD. After realizing the fully charged condition, we measured the PL spectrum by irradiating 532 nm light at 20 W/cm$^2$ and estimated the TOP value. Note that we can neglect irreversible change to the QDs through the irradiation at 532 nm. The spectrum did not show any change for an integration time longer than 120 s, and the obtained TOP value was 70\%, which was much higher than that obtained from a PL spectrum for a fresh QD under the same condition, shown in Fig. \ref{Fig3}(a).  This means that the QD underwent some irreversible change relative to its fresh condition, although  the fully charged condition was not sustained. 
We also measured the PL decay profile for the charged situation. For the excitation, we used a picosecond pulsed laser at 532 nm which may not induce any further irreversible change to the QDs. The obtained profile is displayed in Fig. \ref{Fig5}(c), showing three decay curves with decay times of 1.5$\pm$0.1, 9.9$\pm$0.3, and 131$\pm$6 ns. The red curve shows a fit to the data. We estimated the area for each of the three decay curves, that is a measure of emitted photon number through each decay channel, and estimated the TOP value of 68\%, consistent with the TOP value estimated from the PL spectrum. The results for 405 nm irradiation were similar to those for 455 nm irradiation. TOP values for 405 nm irradiation are plotted in Fig. \ref{Fig4} by green triangles. The TOP value increased from 70\% at 0.5 W/cm$^2$, and reached 100\% at 15 W/cm$^2$. The 405 nm irradiation also induced an irreversible change to the QD, which led to a stationary TOP value of 81\%. 

\begin{figure}[t!]
\centering
\resizebox{0.5\textwidth}{!}{%
  \includegraphics{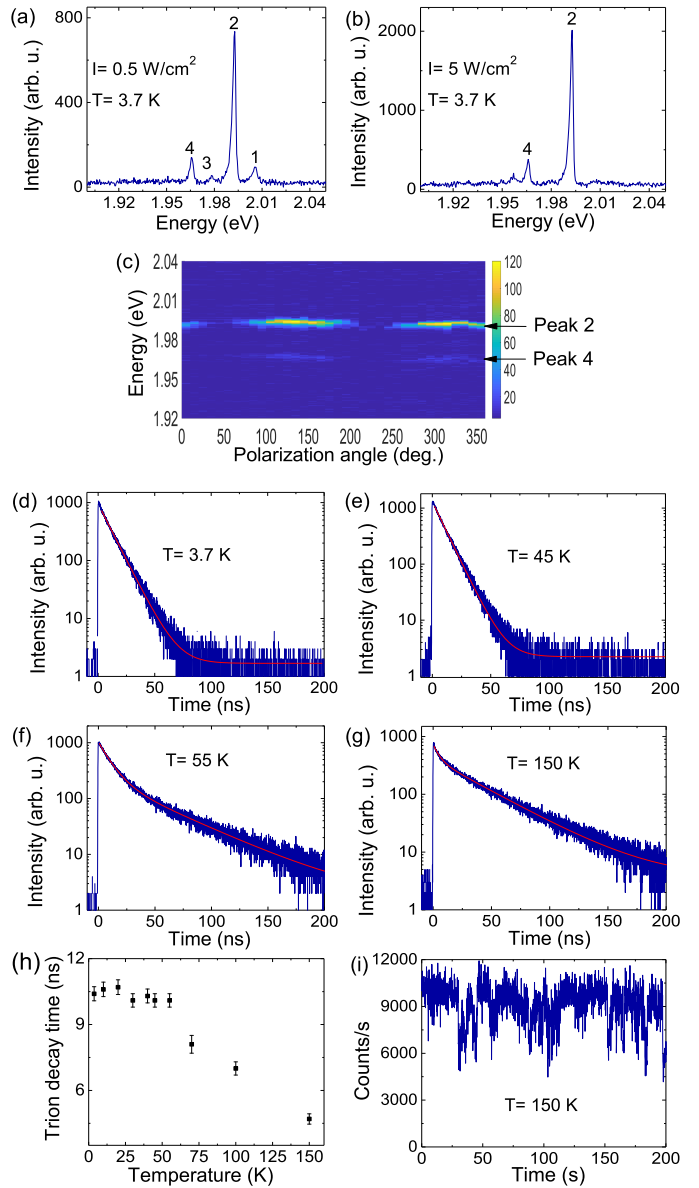}
}
\caption{PL characteristics observed for 355 nm irradiation. (a, b) PL spectra measured for two different irradiation intensities, 0.5 and 5 W/cm$^2$ at 3.7 K. (c) Polarization-resolved PL-spectra under the fully charged condition acquired by rotating an inserted polarizer (analyzer). Peaks 2 and 4 correspond to the trion and its phonon sideband, respectively. Measurements were done using a cw-laser at 532 nm. (d-g) PL decay profile measured by raising the temperature from 3.7 to 150 K plotted in semi-log format. Measurements were done using a ps-pulsed laser at 532 nm after realizing the fully charged condition. (h) Temperature dependence for the trion decay time. Fitted cur  ves are shown in red. Note that decay curves at 3.7 and 45 K are fitted by a single exponential curve. (i) Photon counting behavior for a single QD at 150 K, measured by irradiating 532 nm laser light at an intensity of 20 W/cm$^2$. }
\label{Fig6}
\end{figure}

\subsection{355 nm irradiation}
PL spectra excited at 355 nm at two irradiation intensities with an integration time of 120 s are displayed in Figs. \ref{Fig6}(a, b). One can readily see that, although the exciton peak was seen faintly, the emission was effectively from trion even at an intensity of 0.5 W/cm$^2$ with a TOP value of 86\%. Note that this TOP value is not a stationary value. We measured this spectrum with longer integration times of 180 and 240 s, but the TOP value did not reach a stationary condition. The variation of the measured TOP values is shown as an error bar in Fig. \ref{Fig4}. At an irradiation intensity of 5 W/cm$^2$, only a trion peak with phonon sideband was observed, revealing a fully charged condition. TOP values are plotted in Fig. \ref{Fig4} by blue asterisks. A fully charged condition was realized at higher intensities than 3 W/cm$^2$. In Fig. 6(c), are shown polarization-resolved PL-spectra observed by rotating an inserted polarizer (analyzer) under the fully charged condition. It is readily seen that the PL photons from the trion (peak 2) were linearly polarized. The phonon sideband (peak 4) was also linearly polarized, although the signal was faint. The visibility for the peak 2 was measured to be 85\%. After realizing the fully charged condition, we tested the irreversibility of the irradiation induced process by measuring both PL spectral and temporal responses through 532 nm irradiation, as in the preceding sub-section. The observed PL spectrum exactly reproduced the trion spectrum in Fig. \ref{Fig6}(b), and the temporal response showed a single exponential decay curve with a decay time of 10.4$\pm$0.5 ns as displayed in Fig. \ref{Fig6}(d). The red curve shows the fit by a single exponential function. These observations mean that, different from irradiation by longer wavelengths, the strong irradiation at 355 nm has changed the QD to a permanently charged situation. We should note that this permanent photocharging occurred for all dots we tested. We found that this photocharging lasted as long as the QD was kept at cryogenic temperatures (at least for ten days). We also found that, by raising the temperature of the photocharged QDs to 150 K, the photocharging was removed, and the photocharged QDs returned back to the original situation before irradiation with 355 nm laser light. Figs. \ref{Fig6}(d-g) show temporal characteristics of a photocharged QD by raising the temperature from 3.7 to 150 K. The single exponential behavior was still observed at 45 K, but  a slow decay by exciton started to appear at 55 K. Regarding the decay profiles observed at temperatures higher than 55 K, we analyzed them by fitting with a sum of two exponential curves and derived the trion decay time at each temperature. The obtained trion decay times are plotted in Fig. \ref{Fig6}(h). The trion decay time stayed constant until 50 K and then became faster as the temperature was raised further, revealing an appearance of a nonradiative decay channel and an increase of the nonradiative decay rate via the temperature raise. Fig. \ref{Fig6}(i) shows the photon counting behavior at 150 K measured through 532 nm irradiation at 20 W/cm$^2$.  One can readily see the reappearance of the intermittency, revealing that the intermittency is due to the switching between the exciton and trion states, and the trion QY decreases with the temperature raise due to the increase of nonradiative decay rate. After reaching 150 K, we re-cooled the QD down to 3.7 K, and measured the PL spectrum and decay profile. The PL spectrum was measured through 532 nm irradiation at 20 W/cm$^2$. The observed PL spectrum and temporal profile reproduced essentially the same spectrum and temporal profile, displayed in Fig. \ref{Fig3}(a) and \ref{Fig3}(d), respectively, revealing the revival of the original QD situation before irradiation with 355 nm laser light. 

\section{Discussion}
The observations shown here clearly demonstrate that the charging of QDs, i.e. the occurrence of the trion, can be controlled by irradiating with laser light at cryogenic temperatures, and a degree of photocharging up to 100\% can be manipulated by the irradiation wavelength and intensity. It was also found that, by irradiating a QD with a 355 nm laser light at an intensity higher than 3 W/cm$^2$, the QD can be fully photocharged, lasting in the charged condition permanently as long as the cryogenic temperature was maintained. Regarding the sign of the trion, although we have not identified it experimentally, we assume that the trion is a negatively charged trion, since the measured PL decay time corresponds well to that of negative trions, and the lifetime of the positive trion is expected to be much shorter than the negative trion lifetime \cite{Liu,Rodina}. 

Regarding the photocharging mechanism, although we have not reached the detailed understanding yet, observed photocharging may be qualitatively understood through charge trapping dynamics in the QDs. Regarding the charge trapping for CdSe core-shell QDs, two mechanisms have been discussed so far \cite{Jones,McGuire,Li}. One is based on surface trap sites which are located on the outermost ZnS shell, and the other is based on interfacial trap sites which are located near to the CdSe core. In a PL event, an absorbed photon excites an electron from the valence band to the conduction band, and the created carriers (electron and hole) quickly relax to the band edge, but simultaneously interact with surface charge trap states through a tunnel interaction \cite{Fernee2}. The tunnel interaction can become stronger when the irradiation photon energy is increased from the band edge of the CdSe core (1.75 eV) to the band edge of the outermost ZnS shell (3.54 eV). Such tendency may explain the observations obtained by 455, 405, and 355 nm laser light, photon energy of which are 2.73, 3.06, and 3.50 eV, respectively. 

The irreversible change observed for the irradiation at the three wavelengths may indicate an occurrence of local surface reconstruction induced by the charge/trap-site interaction \cite{Jones}, which may modify the surface sites to realize a more stable situation for the interaction, leading to a deeper trap potential at the site. Such a modified trap potential can become deeper for the 355 nm irradiation whose photon energy is close to the ZnS band edge, and the modified potential can become shallower by lowering the photon energy to 3.06 eV (405 nm) and 2.73 eV (455 nm). The observed monotonic increase of TOP values versus irradiation intensity implies that many excitation cycles are needed to reach a stationary condition for the surface reconstruction. We estimated effective exciton excitation rate to realize a TOP value of 80\% at four excitation wavelengths by estimating photoabsorption cross-section of the QD at each wavelength following a procedure by Leatherdale \it et al. \rm \cite{Leatherdale} using the relative absorbance shown in Fig.\ref{Fig2}(a). Obtained exciton excitation rates were 3.9x10$^6$, 2.3x10$^6$, and 0.2x10$^6$ s$^{-1}$ for irradiation wavelengths of 455, 405, and 355 nm, respectively. The gradual decrease of the excitation rate by increasing the photon energy is reasonably understood from the tendency of the tunnel interaction strength. 

We estimated the trapped charge lifetime for the trap potential modified by 455 and 405 nm irradiation to be shorter than 120 s, because TOP values for these wavelengths showed stationary behavior with an integration time of 120 s. On the other hand, for the 355 nm irradiation, the trapped carrier lifetime was effectively infinite for irradiations higher than 3 W/cm$^2$. Regarding the effective potential depth of the modified trap potential realizing a permanently charged situation, it could be estimated to be around 4 meV, since the permanently charged condition was sustained up to 50 K. It should be noted that such a modified trap potential, i.e. surface reconstruction, can be removed by raising the temperature, since observations showed that the charged QDs return back to the original QD situation before charging by raising the temperature to 150 K. We should also note that it is known that hole traps constitute the majority surface trap sites \cite{Lifshitz,Abdellah}, consistent with the assumption that the observed trion is a negatively charged trion. 

Regarding the 532 nm irradiation, the observed TOP behavior was different from those for other three irradiations. TOP value did not reach 100\%, but saturated at 90\%. Strong irradiation did not induce any irreversible change observed for the shorter wavelength irradiations. Moreover, estimated exciton excitation rate to realize 80\% TOP was 1.9x10$^6$ s$^{-1}$, which is smaller than the values obtained for shorter wavelength (higher photon energy) irradiation at 455 and 405 nm. These observations suggest that the charge trapping mechanism for the 532 nm irradiation is not due to the surface trap sites. We suspect that the charge trapping may be due to the interfacial core trap sites which are located near to the CdSe core and for which reconstruction may not occur.

Lastly, it should be noted that although we have thus qualitatively explained the observations through charge trapping dynamics, the explanation would still be a speculation. To fully understand the mechanism, we expect further systematic investigations for the QDs by manipulating the QD conditions, such as shell thickness or types of surface ligands.

\section{Conclusion}

We have investigated the photocharging behavior for CdSe QDs systematically, leading to a technique to control the photocharging of the CdSe QDs at cryogenic temperatures. Most importantly, we have established a method to create permanently charged QDs with trion emissions as long as the cryogenic temperatures are maintained. The present method may open a new route to apply the QDs for quantum photonics, since the trions at cryogenic temperatures give much more preferable PL characteristics than those for neutral excitons. It should be noted also that the present technique has been demonstrated using a QD/ONF hybrid system which can be readily integrated into fiber networks. The technique may be directly applicable to quantum photonic devices, such as fiber-in-line single photon generators. The present results may also trigger research works into more detailed understanding of the photocharging mechanism.

\section{Acknowledgements}
The authors thank Mark Sadgrove for his careful reading and critical comments to the manuscript. This work was supported by the Japan Science and Technology Agency (JST) through Strategic Innovation Program (Grant No. JPMJSV0918).

\end{document}